\documentclass[useAMS,usenatbib,usegraphicx]{mn2e}
 
\title[Observational Evidence for Gas Accretion]{Gas Accretion as a Dominant Formation Mode in Massive Galaxies from the GOODS NICMOS Survey}
\author[Conselice et al. ]{Christopher J. Conselice$^{1}$\thanks{E-mail:
conselice@nottingham.ac.uk}, Alice Mortlock$^{1}$, Asa F.L. Bluck$^{1,2}$, Ruth Gr\"utzbauch$^{1}$, \newauthor Kenneth Duncan$^{1}$\\
$^{1}$University of Nottingham, School of Physics \& Astronomy, Nottingham, NG7 2RD UK \\
$^{2}$Gemini Observatory }

\def\solm{M$_{\odot}\,$}

\def\solm{M$_{\odot}\,$}

\def\mass{$10^{11}$ M$_{\odot}\,$}

\def\casgm20{CAS-G-M$_{20}\,$}
\def\m20{M$_{20}\,$}
\def\ms{M$_{*}\,$}
\def\mg{M$_{\rm g}\,$}

\begin{document}

\date{Accepted ; Received ; in original form}
\pagerange{\pageref{firstpage}--\pageref{lastpage}} \pubyear{2002}

\maketitle

\label{firstpage}

\begin{abstract}

The ability to resolve all processes which drive galaxy formation is one of
the most fundamental goals in extragalactic astronomy.  While star formation
rates and the merger history are now measured with increasingly high
certainty, the role of gas accretion from the intergalactic medium 
in supplying gas for star formation still
remains largely unknown.   We present in this paper indirect
evidence for the accretion of gas into massive galaxies with initial stellar
masses M$_{*} > 10^{11}$~\solm and following the same merger
adjusted co-moving number density
at lower redshifts during the epoch $1.5 < z < 3$, using results
from the GOODS NICMOS Survey (GNS).    Our method utilises 
the observed star formation rates of these massive galaxies based on UV
and far-infrared observations,  and the amount of stellar and gas mass added due to observed
major and minor mergers to calculate the evolution of stellar mass in these 
systems.  We show that the measured gas mass fractions of these
massive galaxies are inconsistent with
the observed star formation history for the same galaxy population.  We further
demonstrate that this additional gas mass cannot be accounted for by cold gas 
delivered through minor and 
major mergers.   We also consider the effects of gas outflows and gas recycling
due to stellar evolution in these calculations.
 We argue that to sustain star formation at the observed rates 
there must be additional methods  for increasing the cold
gas mass, and that the 
likeliest method for establishing this supply of
gas is by accretion from the intergalactic medium.  
We calculate that the average gas mass accretion rate into these massive 
galaxies between $1.5 < z < 3.0$, is 
$\dot{M} = 96\pm19$ \solm yr$^{-1}$ after accounting for outflowing gas.   
This is similar to what is predicted in
detailed simulations of galaxy formation.   We show
that during this epoch, and for these very massive galaxies, 49$\pm$20\% of 
baryonic mass assembly is a result of gas accretion and unresolved mergers, 
while the remaining $\sim 25\pm10$\% 
is put into place through existing stars from mergers, with the
remainder is gas brought in with these mergers. 
However, 66$\pm20$\% of all star formation in this
epoch is the result of gas accretion.  This reveals
that for the most massive galaxies at $1.5 < z < 3$ gas accretion 
is the dominant method for instigating new stellar mass assembly.

\end{abstract}


\section{Introduction}

Both observations and theoretical models now overwhelmingly suggest that 
galaxies have evolved from an early population 
dominated by lower mass systems undergoing significant star formation 
in the early universe, to the large and relatively passive galaxies 
that we find today (e.g., Conselice 2006; Bouwens et al. 2010). How 
this transformation occurs, that is how we get from 
young low mass galaxies to the large massive galaxies we see in 
today's universe,  is a highly debated topic.  
Essentially, we want to answer the question - how 
do galaxies assemble their stellar masses?  The answer to this question 
will have profound implications for both the physics of galaxy
formation and for understanding properties of the universe itself.

Historically, it was once thought that galaxies formed in a manner 
similar to stars through a collapse of gas that later, through some 
process, underwent intense star formation.  In this paradigm the total baryonic 
mass of a galaxy does not change significantly with time, and its stellar 
mass evolves  by rapidly converting  gas into stars.   
However, it is clear that galaxies must form in a 
hierarchical way through mergers and accretion of material from 
the intergalactic medium.  This view has supporting evidence from both 
observations of the stellar mass functions of galaxies (e.g., Bundy et al. 2006; 
Mortlock et al. 2011), direct observations of mergers in the distant 
universe (e.g., Conselice et al. 2003, 2008; Bluck et al. 2009, 2012), 
as well as through the evolution of galaxy 
sizes (e.g., Ferguson et al. 2004; Trujillo et al. 2007; Buitrago et al. 2008;
Weinzirl et al. 2011).    
Evidence from internal kinematics also suggests that galaxy formation is driven 
at least in part by the accumulation of gas from the intergalactic medium falling 
onto a galaxy (e.g., Keres et al. 2005; Genzel et al. 2008; 
Dekel et al. 2009a,b; Bournaud et al. 2011).

While it is now generally accepted that galaxy formation is a hierarchical 
process driven by the accumulation of stars and gas located outside of a galaxy after
its initial formation, the 
details of this assembly remain largely unknown.  Cold gas infall into galaxies, 
otherwise known as `cold gas accretion' is theorised to 
be an important aspect in the process of massive galaxy formation (e.g.,  White \& Frenk 
1991; Birnboim \& Dekel 2003; Keres et al. 2005,
Dekel et al. 2009a,b), and even perhaps the dominant method by which galaxies assemble 
their mass.   However, there is little to no evidence for gas accretion onto 
galaxies at high redshift currently (e.g., Steidel et al. 2010), although
some claims are appearing at lower redshifts (e.g., Rauch et al. 2011).    
This is primarily due to the 
difficultly of observing this process since the covering fraction of 
accreting cold gas is likely not high (e.g., Faucher-Giguere, Keres \& Ma 2011; 
Kimm et al. 2011; Fumagalli et al. 2011), 
nor would this gas be easily detected in, for example, absorption
(e.g., Weiner et al. 2009; Giavalisco et al. 2011).

In this paper we develop a new approach to this problem by analyzing 
the evolution of the 
massive galaxy population in terms of the history of its baryonic 
assembly.  We examine how the stellar mass of a massive galaxy is built up 
during $1.5 < z < 3$ through various galaxy formation processes observed 
within a unique sample of M$_{*} >$ \mass galaxies  taken 
from the GOODS NICMOS Survey (GNS) (Conselice et al. 2011).   By examining
the addition of stellar mass due to major and minor mergers (Bluck et al. 
2009; 2012), and the observed star formation history (Bauer et al. 2011)
and resulting stellar mass evolution using stellar synthesis modeling. we 
provide through this method circumstantial evidence for gas inflow, or
gas imported in through extremely minor mergers, as an 
important process in galaxy assembly.   

After comparing the amount of accreted
mass to the mass assembled through merging, we conclude that gas accretion 
is a dominant process for massive galaxy assembly at this epoch.  
We further discuss the comparison with theory, and
describe some of the implications of our results, including how our
work relates to the G-dwarf problem (e.g., Larson 1974), and the
rapid gas depletion time-scales in galaxies.  

This paper is organised as follows: \S 2 includes a discussion of the 
data sources we use in this paper, and the sample selection, \S 3 is a 
description of our baryonic mass assembly analysis,
\S 4 presents our arguments for gas accretion and \S 5 is our summary.   
We use a standard cosmology of H$_{0} = 70$ km s$^{-1}$ Mpc$^{-1}$, and 
$\Omega_{\rm m} = 1 - \Omega_{\lambda}$ = 0.3 throughout.

\section{Data, Data Products and Galaxy Selection}

\subsection{Data Sources}

The data and methods we use in this paper originate from the GOODS NICMOS
Survey (GNS), and are described in detail in Conselice et al.
(2011), Mortlock et al. (2011), Bluck et al. (2009, 2012) and Bauer et al.
(2011).  The primary galaxies and parent sample we examine are 
81 massive galaxies with stellar masses
M$_{*} > $ \mass at redshifts $1.5 < z < 3$.  These were selected through
a variety of colour selections, include Distant Red Galaxies (DRGs), IRAC
Extremely Red Objects (IEROs) and the BzK selected systems (Conselice et al. 2011).
We also utilise photometric redshifts which are described in detail in
Gr\"utzbauch et al. (2011a,b).  

\begin{figure}
 \vbox to 120mm{
\includegraphics[angle=0, width=90mm]{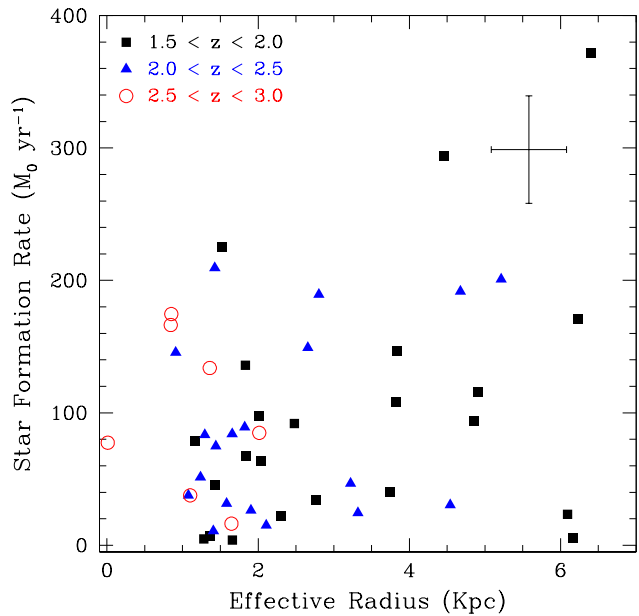}
 \caption{The relationship between the star formation rate and effective radius, 
R$_{\rm e}$, divided into three redshift ranges between $1.5 < z < 3$ for 
galaxies with stellar masses M$_{*} > 10^{11}$ \solm.}
} \label{sample-figure}
\end{figure}
  
To calculate stellar masses we use
a Bayesian method to fit spectral energy distributions based on
various star formation models to the galaxy photometry.  The distribution of the 
resulting stellar masses gives
us an error in their measurement, and we take the peak value of the distribution as
a measure of our stellar mass. A Salpeter IMF was used in these calculations.  
Further details about our sample selected are included in Conselice
et al. (2011) where we refer interested readers for more details.

The galaxy effective radii which we use in this study to measure star formation 
rate surface densities, originate from GALFIT fits to our NICMOS imaging, 
taken
from the methods and catalog of Buitrago et al. (2008, 2011).  These are single
S{\'e}rsic profile fits to the light profile in which we retrieve both a S{\'e}rsic
index, $n$, as well as the effective radius, R$_{\rm e}$.  These sizes are 
measured solely using the GOODS NICMOS Survey H$_{160}$-band imaging.  
Detailed simulations show that we are able to retrieve these parameters
given the observing conditions of our sample (Buitrago et al. 2011).

We also utilise results from previously published GNS studies which 
describe the evolution of this sample
in terms of the star formation rate (Bauer et al. 2011) and the merger history
(Bluck et al. 2009; 2012) as a function of stellar mass.  We give brief 
summaries of these analyses in 
reference to our current study throughout this paper.  The star formation rates
we use for our galaxies are measured through rest-frame UV and Spitzer 24 $\mu$m
measurements, which are consistent with each other (Hilton et al. 2012).  
For the UV star formation 
rates, which are used for most of the
analysis here, we utilise ACS z-band imaging to obtain a rest-frame UV flux
measurement which is converted to a star formation rate after applying
a k-correction (Bauer
et al. 2011).  We further measure the UV slope for each galaxy using the UV
colour to correct for dust extinction.

\subsection{Galaxy Selection}

Our galaxy sample is initially chosen simply as all systems within our sample which
have stellar masses of M$_{*} > 10^{11}$ \solm at redshifts of 
$2.75 < z < 3$.    We select all galaxies within these limits, including 
passive galaxies or star forming galaxies.  Our range of star formation measures
for our massive galaxies include systems that would be identified by colours
as passive, but which still have measurable star formation rates, albeit not
as high as more active systems.  In fact, the star formation rates measured
for our sample range from less than 10 \solm year$^{-1}$ to over 
300 \solm year$^{-1}$, with a small fraction of 5\% of systems having a star
formation rate too low to measure.  

We include all of the galaxies in our
co-moving number density selection as the
star formation rate and gas mass fractions we use are an average within this 
population.  The resulting conclusions are therefore valid for the `average' 
galaxy within our selection.   We assume that our sample of galaxies 
represents different phases of a 
massive galaxy's life-span throughout the redshifts $1.5 < z < 3$.
For example, the fraction of 
galaxies which are passive represents the fraction of time a typical galaxy in our 
sample is not undergoing star formation.    Our sample of galaxies all have
very similar stellar masses and environments, suggesting that these systems
should have similar formation histories (e.g., Gr\"utzbauch et al. 2011a,b).
The result of this is that we can obtain the average amount of stellar mass
added to this sample selection due to star formation and merging without
having to worry about biases from different evolutionary phases.


One issue that we have to account for is that the initial selection of
M$_{*} > 10^{11}$ \solm at redshifts of $2.75 < z < 3$ cannot be traced
at a constant mass limit.
In Conselice et al. (2011) and Mortlock et al. (2011) we discuss the
stellar mass evolution for our sample.  By examining our mass selection used
in this paper we find that the number density of galaxies selected by the
criterion M$_{*} > 10^{11}$ \solm grows by a factor of between two and four from
redshifts $z = 3$ to $z = 1.5$.  We therefore examine in this paper
our galaxy sample at a constant merger adjusted co-moving number density, taking
account of both mergers and star formation which changes the stellar masses of
our initially selected galaxies and those below our initial mass limit. 

We find that starting with a mass range of log M$_{*} > 11$
at $2.5 < z < 3$ due to changes in star formation and merging (Bluck
et al. 2012) this mass limit grows to log M$_{*} = 11.2$ at 
redshifts $2.0 < z < 2.5$ and log M$_{*} = 11.3$ at $1.5 < z < 2$.  Overall
on average a galaxy will undergo around a single major merger, thereby
decreasing the number of systems in our initial selection
during this epoch. We account for this by tracing at a resulting 
decreased number density to follow the same original galaxies.

Figure~1 shows a summary plot of our data and massive galaxy 
sample. Plotted are the star formation
rates vs. the effective radii for each galaxy in our sample, divided into different
redshift bins.  The average star formation rate for our sample is roughly constant
with redshift (Bauer et al. 2011), and there is a slight evolution in the sizes
of these galaxies, such that they are growing with time (e.g., Buitrago et al. 2008).

\section{Evolution of Galaxy Mass}

\begin{figure}
 \vbox to 120mm{
\includegraphics[angle=0, width=90mm]{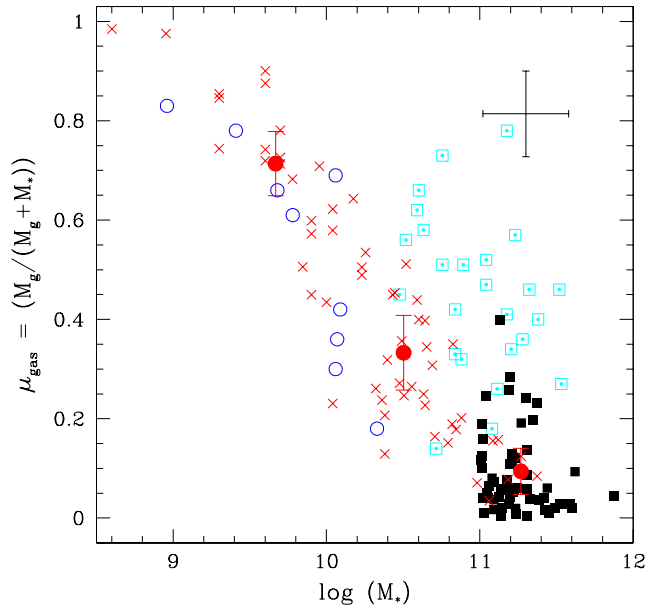}
 \caption{The relation between stellar mass (M$_{*}$) and the 
gas mass fraction $\mu_{\rm gas}$, which is the ratio of the cold gas
mass divided by the stellar+gas mass.  The red crosses are for $z > 2$
galaxies from Erb et al. (2006) and the blue circles are from Mannucci
et al. (2009).  The solid black boxes are from the GNS (this work).  The error
bars show the averages and the 1 $\sigma$ dispersion as a function of
stellar mass.  The cyan points with square boxes are the CO measurements of starbursting
galaxies at similar redshifts from Daddi 
et al. (2010) and Tacconi et al. (2010).}
} \label{sample-figure}
\end{figure}

\subsection{Galaxy Gaseous Mass at $z > 1.5$}

The evolution of the cold gas mass within galaxies is an
important aspect for understanding galaxy evolution, yet our 
observations of 
this quantity are still very basic and are prone to systematic errors.  
Furthermore,  all of these measurements
are indirect, even when using CO as a proxy.  We present in this
section arguments for how to quantify gas masses in our galaxy sample
based on the inverse Schmidt-Kennicutt relation (e.g., Kennicutt 1998;
Erb et al. 2006;
Daddi et al. 2010; Boquien et al. 2012) using our measured
SFRs and sizes (Bauer et al. 2011, Buitrago et al. 2011; \S 2).

\begin{figure}
 \vbox to 140mm{
\includegraphics[angle=0, width=90mm]{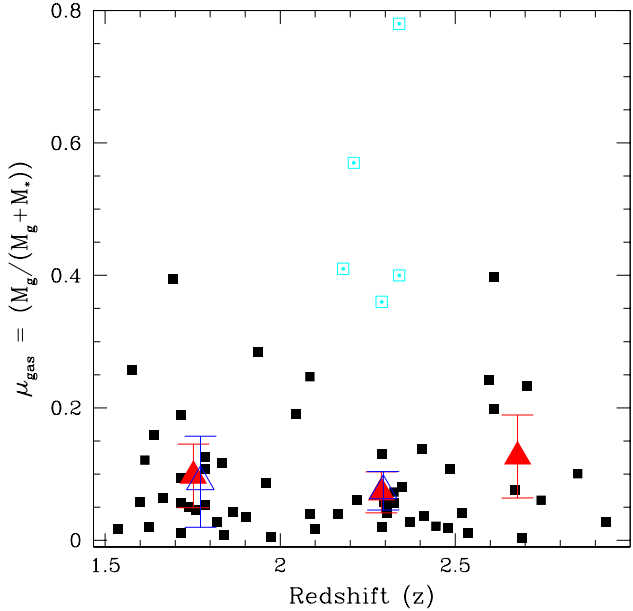}
 \caption{The relation between the gas mass fraction and redshift for galaxies
with stellar masses M$_{*} >$ \mass.  The error bars show the 1 $\sigma$ scatter for
the average values which are plotted as large solid triangles.  The red triangles are
for this method when examining galaxies at a stellar mass cut off of log M$_{*} > 11$,
and the open blue triangles are when holding at a constant merger adjusted
co-moving number density to
account for more galaxies entering our mass selection over time.    The cyan points
surrounded by boxes are the galaxies from Daddi et al. (2010) and Tacconi et al. (2010) which
meet our stellar mass criteria.}
\vspace{2cm}
} \label{sample-figure}
\end{figure}

Obtaining a measure of the cold gas mass in galaxies is challenging, and 
there are several
attempts to measure this at high redshift (e.g., Erb et al. 2006; Mannucci et 
al. 2009;  Daddi 
et al. 2010), yet these methods contain several important limitations 
which must be considered.  These inferred cold gas mass fractions
are measured through the inverse Schmidt-Kennicutt
relation, which relates the cold gas mass surface density to the star formation rate 
surface density (e.g., Kennicutt 1998).  This relation, for nearby galaxies,
has been studied in detail
in recent papers such as Bigiel et al. (2011) who find a significant
scatter, but still a correlation, between these two quantities using
molecular gas and star formation measurements.  Much of this scatter is
however proposed by Krumholz et al. (2012)  to be due to observational
projection effects,
and that there is a universal star formation-gas surface density law that applies
at all redshifts (see also Narayanan et al. 2011).  There is also evidence 
that at least two star formation laws
may apply at higher redshifts, one for disk-like galaxies, and another for 
starbursts (e.g., Bouche et al. 2007; Daddi et al. 2010; Genzel et al. 2010).

Most high-z observations, including ones based on CO detections, 
argue that the cold gas mass fraction, M$_{\rm g}$/M$_{*}$ is between
10-50\% for galaxies more massive than M$_{*} = 10^{10}$ \solm (e.g.,
Genzel et al. 2010; Daddi et al. 2010; Figure~2).  
This cold gas mass fraction tends to rise for lower mass
galaxies (e.g., Erb et al. 2006; Mannucci et al. 2009). There is 
also evidence that the efficiency
of star formation within starbursts, such as ULIRGs and sub-mm galaxies 
is more efficient than that given by the standard
Schmidt-Kennicutt law, resulting in lower derived gas mass fractions at a
measured
star formation rate surface density (e.g., Daddi et al. 2010; Boquien 
et al. 2012).

We calculate gas masses for our massive M$_{*} > 10^{11}$~\solm sample, and 
derive the cold gas mass fraction
using a form of the global Schmidt-Kennicutt law calibrated for nearby
star forming galaxies.     The relation we use is:

\begin{equation}
\Sigma_{\rm SFR} = (2.5\pm0.7) \times 10^{-4} \left(\frac{\Sigma_{\rm gas}}{1\, {\rm M_{\odot}} {\rm pc^{-2}}} \right)^{1.4 \pm 0.15} {\rm M_{\odot}}\,  {\rm yr^{-1}} {\rm kpc^{-2}},
\end{equation}

\noindent  where $\Sigma_{\rm SFR}$ is the surface density of star 
formation, and $\Sigma_{\rm gas}$ is the surface density of cold gas.  
We assume in this calculation that the star formation follows the distribution 
of H-band light, which is found to be the case (Ownsworth et al. 2012). In any case
the exact form of the profile in star formation and light are not important as long
as the effective or half-light radius,  R$_{\rm e}$, are the same. 
We calculate the star formation rate surface density within each galaxy
based on the effective radius, R$_{\rm e}$ (\S 2), and half of the measured total star formation 
rate.   From this we obtain the gas mass surface density using eq. (1) and we then 
calculate the total gas masses in our systems based on this.

In Figure~2 we plot the gas mass divided by total baryonic mass (M$_{*}$ + M$_{\rm g}$) for our
sample of massive galaxies.  We also show in Figure~2 comparison values from
Erb et al. (2006) who study galaxies at similar redshifts, but at lower masses, and Mannucci et
al (2009) who study similar star forming galaxies at $z > 2$.   We also plot on
Figure~2 and Figure~3 the direct measurements of cold gas masses from Tacconi
et al. (2010) and Daddi et al. (2010) who both find gas fractions towards the upper
end of our stellar mass selected sample.  This is not unexpected given that these
two studies measure gas masses for galaxies with the highest star formation rates
at these redshifts.

Figure~2 shows
that the galaxies in our sample, as plotted by the black boxes, have gas total mass fractions which
are around 10\%, with some scatter.  However, our values are lower than the gas total mass fraction
found for lower stellar mass systems, and extend the trend found in Erb et al. (2006) and
Mannucci et al. (2009), such that the highest mass systems have the lowest relative 
gas fractions.

To determine how much more stellar mass
these galaxies could acquire from the existing cold gas in star formation events,
we show in Figure~3 the gas mass fraction as a function of redshift.  The average of the ratio of the gas 
mass to stellar mass at the three redshift ranges we examine, and the 1$\sigma$ 
dispersions of these ratios are at $z \sim 1.75$ - $f_{\rm g} = 0.13$, $\sigma = 0.15$, $z \sim 2.29$ -
$f_{\rm g} = 0.08, \sigma = 0.08$ and at $z \sim 2.68$ - $f_{\rm g} = 0.17, \sigma=0.19$.  

We compute these averages when we hold our selection at a constant merger
adjusted co-moving number density as 
the open blue triangles on Figure~3.  We thus
find, as others have (Mannucci et al. 2009), that the gas mass
fraction is roughly constant with redshift for both
a constant mass selection, and using a constant merger adjusted 
co-moving number density 
selection.  This
is also found by comparing derivations of gas mass fractions at different
redshifts for lower mass galaxies (e.g., Erb et al. 2006; Mannucci et al. 2009).

It has been noted in previous studies that the gas masses and depletion times
scales are not consistent with the star formation observed in these systems 
(e.g., Genzel et al. 2010).  Figure~4 displays
the depletion time-scale, the gas mass divided by the star formation rate ($\psi$), 
i.e., M$_{\rm g}$/$\psi$ -- showing that nearly all galaxies in our parent sample
would be depleted of gas in less than 0.5 Gyr assuming 100\% efficiency of
star formation. This is much less than the time within the 
redshift interval we examine in this paper.   While some of these galaxies
may become red/passive systems very quickly, the average galaxy within our
co-moving number density selection still contains a high star formation rate.
This suggests that something is replenishing the gas within the average galaxy 
within our selection over time.

\begin{figure}
 \vbox to 120mm{
\includegraphics[angle=0, width=90mm]{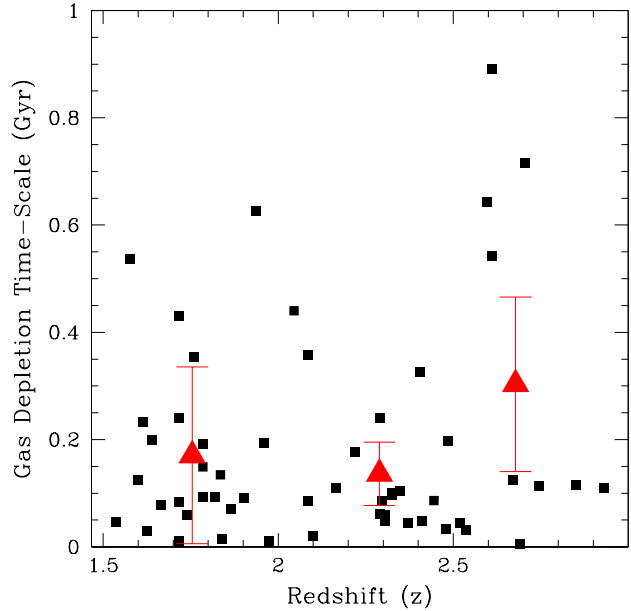}
 \caption{The gas depletion time scale shown as a function of redshift for 
each galaxy within our sample.  The time-scale is in Gyr, and the triangles show
the average values at different redshift ranges, and the errors bars are the
1$\sigma$ scatter.  }
} \label{sample-figure}
\end{figure}

\subsection{The Evolution of Stellar Mass from Star Formation}

Galaxy stellar mass is a measure of both the amount of gas which has been converted
into stars over time, as well as the amount of mass which has been
accreted into the galaxy from previously existing galaxies. We are able to 
trace both of these properties for the most massive galaxies in the
universe up to $z \sim 3$ (e.g., Conselice et al. 2011; Mortlock et al. 2011;
Bauer et al. 2011; Bluck et al. 2012).

The reason we can do this now is that we have constructed a complete
sample of massive galaxies at redshifts $z = 1.5$ to $z = 3$ which have stellar masses in excess
of M$_{*} = 10^{11}$ \solm (Conselice et al. 2011) that have measured minor+major
merger measurements (Bluck et al. 2012).   We examine
galaxies within this selection at $2.75 < z < 3$ and follow at lower redshifts
the same systems by observing galaxies at a constant co-moving merger
adjusted number density.
Using this method, we are ensuring that we are examining
similar galaxies at different redshifts (\S 2.2). 

In previous work, we carried out detailed calculations of the merger and
star formation histories for a sample of 81 massive galaxies
at $1.5 < z < 3$ (Bauer et al. 2011; Bluck et al. 2012) which we discuss
here.  We find that the average star formation 
rate for galaxies with stellar mass, M$_{*} > 10^{11}$ \solm is $<\psi> =$ 
103$\pm$8 \solm yr$^{-1}$ (Bauer et al. 2011), with a dispersion of
75 \solm yr$^{-1}$.  We find very similar star formation rates for the
same sample when we examine their {\em Herschel} far-infrared spectral energy
distributions (Hilton et al. 2012).  We furthermore measure the
star formation rate for galaxies at a constant co-moving merger adjusted
number density at lower redshifts, from our initial sample. 
When we do
this we find that the star formation rate remains at near 100 \solm yr$^{-1}$, with
a slight decline from 110 \solm yr$^{-1}$ to 100 \solm yr$^{-1}$ over the interval
$1.5 < z < 3$.  We use this star formation rate history at a constant merger
adjusted co-moving 
number density when calculating the formation history of our initial
sample of M$_{*} > 10^{11}$ galaxies.


The observations we use in this paper are measured from redshifts $z = 3$ to $z = 1.5$, 
a time period of roughly  2.16 Gyrs.  We calculate how much stellar mass is
created during
this redshift interval by integrating the star formation rate, $(\int \psi \times \delta t$).
We also account for mass
loss due to stellar evolution processes which we denote as M$_{\rm *, recy}$.  
When we do this, we obtain the average amount of stellar mass created
per galaxy over this redshift interval, the value of which is: 
$(\int \psi \times \delta t$) $\sim (2.24\pm0.34) \times 10^{11}$ \solm.
This is roughly the average amount of stellar mass originally in our sample of 
galaxies at our highest redshift, $z \sim 2.75 - 3$, which we denote 
as M$_{*}$(t=0) = M$_{*}(0)$. 

The term (M$_{\rm *,recy}$) accounts for  
the amount of stellar matter brought back into the interstellar medium of
these galaxies due to SN, and other stellar evolutionary events using
Bruzual and Charlot (2003) models.  Using a constant star formation
model, which matches the observations, we obtain the net amount of 
stellar mass created after considering the amount brought back
into the ISM through these process, using a Salpeter IMF.   We find
that on average this amounts to 18\% of the stellar mass in
star formation is brought back to the ISM of each galaxy. An integral of this is
done over our time-span from $1.5 < z < 3$ and the resulting amount of
stellar mass converted back into gas is denoted as M$_{\rm *, recy}$.

Thus we can measure the
ratio of the net stellar mass created in star formation within these galaxies
divided by the original stellar mass within these systems, or,

$$\frac{(\int \psi \times \delta~t) -{\rm M_{*,recy}}}{\rm M_{*}(0)}~=~1.00\pm0.15.$$  

\noindent This is then the amount of gas mass converted into stars due to
the star formation process which remains after stellar evolutionary
effects have been accounted for at a constant merger adjusted co-moving
number density. 

\subsection{The Evolution of Stellar Mass from Minor+Major Mergers}


When calculating the accretion of stellar mass into our sample of
galaxies, we have to consider both major
and minor mergers, as another major route for galaxies
obtaining stellar mass.
The amount of stellar mass added to a galaxy due to the merger
process is given by the integral over the merger history, based on the
fraction of galaxies merging, and the time-scale for mergers (e.g.,
Conselice et al. 2003; Conselice 2009; Bluck et al. 2009; 2012).  We carry 
out this integration 
using the observed merger history and time-scale for mergers measured
directly from our sample (e.g., Bluck et al. 2009; 2012).   

As detailed in Bluck et al. (2012) the merger fraction
can be parameterised as a function of both stellar mass and redshift.  Bluck et 
al. (2012) find that at 2.3 $<$ z $<$ 3  the merger fraction dependence on 
stellar mass for galaxies within our stellar mass range is given by:


\begin{equation}
f_{m}(M_{*}) = (0.28\rm{+/-}0.17) \times \delta {\rm log} (M_{*})^{0.91+/-0.35}
\end{equation}

\noindent and at 1.7 $<$ z $<$ 2.3 Bluck et al. (2012) find the merger fraction for the same co-moving
number density of galaxies is:

\begin{equation}
f_{m}(M_{*}) = (0.16\rm{+/-}0.11) \times \delta {\rm log} (M_{*})^{1.18+/-0.35} .
\end{equation}

\noindent The total amount of stellar mass accreted is then a double integral over the 
redshift range 
of interest and over the stellar masses which we probe, which for the GNS is sensitive down
to M$_{*} = 10^{9.5}$ \solm, can be written as,

\begin{equation}
{\rm M_{*, M}} = \int_{z_{1}}^{z_{2}} \int_{M_{1}}^{M_{2}} M_{*} \times \frac{f_{m}(z,M_{*})}{\tau_{\rm m}} dM_{*} dz.
\end{equation}

\noindent Where $\tau_{\rm m}$ is the merger time-scale, which depends on the
stellar mass ratio of the merging pair (Bluck et al. 2012).   The value of $\tau_{\rm m}$
ranges from 0.4 Gyr for a 3:1 stellar mass ratio merger, 1 Gyr for a 9:1 stellar
mass ratio merger based on N-body/hydrodynamical simulations, while even higher mass ratios can be even longer
with a logarithmic dependence on mass ratio (see Bluck et al. 2012 for details).    
Our total integration gives a value of 

$$\frac{{\rm M}_{\rm *,M}}{\rm M_{*}(0)} = 0.56\pm0.15.$$ 

\noindent This is the amount of stellar mass added due to both major and minor mergers
for systems with stellar mass ratios down to 1:100 for the average system in our sample. 
After
taking into account baryonic mass, this reduces to mass ratios of 1:20.  We cannot
extrapolate this to lower mass ratios easily as we have no data on the lower mass
merger
fractions, but down to log M$_{*} = 8.5$ we would add a small
amount of mass, but the merger time-scales for these extremely
low mass mergers is likely much too long to have merged by $z = 1.5$ 
(e.g,. Bluck et al. 2012).  

We integrate the amount of gas mass added due to mergers in a similar way, based on the
ratio of stellar to gas mass, based on the data shown in Figure~2.  We fit an empirical
relationship between the gas mass fraction $\mu_{\rm gas}$ and the stellar mass, finding

$$\mu_{\rm gas} = -0.35\times {\rm log\, M_{*}} + 4.13.$$

\noindent This relation allows  us to compute the
total amount of gas, and the total amount of stellar mass accreted from these minor 
mergers.  We show in Figure~5 the relative amounts of gaseous and stellar mass 
as a function of the stellar mass of the merging galaxy.    This shows that the 
contribution of stellar mass to an average massive galaxy in our sample from
merging systems is highest around M$_{*} = 10^{10.8}$~\solm and declines at lower and 
higher masses.  The
gas mass accreted however is quite low at the highest mass galaxies, but increases 
at lower masses and stays relatively constant at values M$_{*} < 10^{10.5}$~\solm.

Using this, we calculate how much gas mass is brought into our galaxies due to 
these minor mergers. We calculate that the average stellar mass weighted gas  
mass fraction is 
f$_{\rm g} \sim 1.03$ down to M$_{*}$ = $10^{9.5}$~\solm.  From this, we find
M$_{\rm g, M}$/M$_{*}(0)$ = M$_{\rm *,M}$/M$_{*}(0)$ 
$\times$ f$_{\rm g}$ = 0.57$\pm0.15$.   We assume here that the relation between
gas density and stellar mass is constant over the redshift interval $1.5 < z < 3$.  
We find this to be the case for our massive galaxies, and if the gas within the
lower mass galaxies declined at lower redshifts, we would only need more gas
accretion to account for the observed star formation, making our measurement 
a lower limit in this sense.  
 This gas contribution
from the minor mergers is our key new measurement that allows us to later
constrain how much gas is coming from the intergalactic medium to form galaxies.

\begin{figure}
 \vbox to 120mm{
\includegraphics[angle=0, width=90mm]{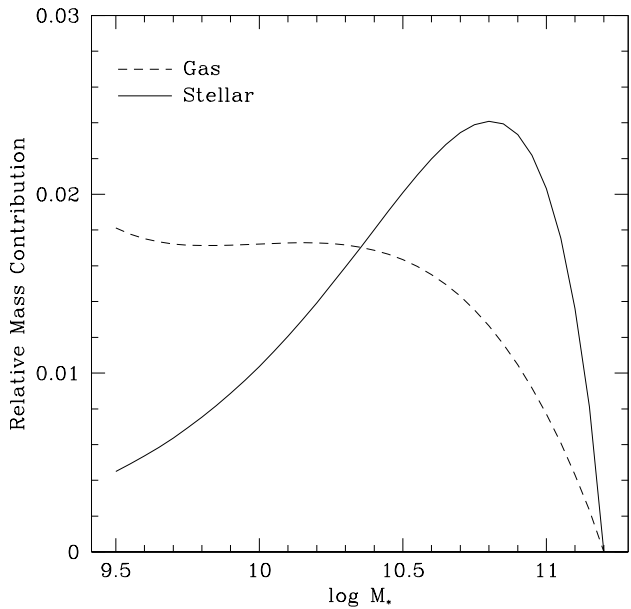}
 \caption{The relative amount of gaseous and stellar matter accreted from mergers of various
masses between redshifts $1.5 < z < 3$.  The x-axis is the stellar mass of galaxies merging with
our typical average galaxy from our sample with M$_{*} = 10^{11.2}$ \solm.    The y-axis shows
the relative contribution to the total for galaxies at masses separated by 0.05 dex.  The solid
line shows the stellar mass contribution and the dashed line the gaseous contribution.}
} \label{sample-figure}
\end{figure}

\section{Gas Accretion}

\subsection{Formulation of Problem}

A galaxy has several mass components that interlink with each other. 
These are the stellar mass (\ms), cold gas mass (\mg) and dark matter mass.  
For the purposes of this paper we consider that the gas mass is
cold, and thus any hot gas or gas that cannot be converted into stars
quickly
is not included here.  Hot gas is present within massive galaxies at least 
in the local universe, and
it remains possible that some of this hot gas is cooling to form stars within 
the systems we observe here.  What we are interested in is the amount of 
new cold gas which arises from either gas accretion from the intergalactic 
medium or from hot gas within the halos of these galaxies cooling, which 
remains a theoretical possibility 
(e.g. Keres et al. 2005;  Ocvirk et al. 2008).   Our method is ultimately not
able to distinguish between these possibilities.

The total baryonic mass (M$_{\rm bary}$) of a galaxy can be written as:

\begin{equation}
{\rm M}_{\rm bary} = {\rm M_{*}} + {\rm M_{g}} 
\end{equation}

\noindent where these terms are all expressible as a function of time. We
first calculate using these terms the amount of stellar mass within a
galaxy as a function of time M$_{*}(t)$.  As the stellar mass in a galaxy will 
not significantly
decrease, the value of M$_{*}$ can in general only increase. This increase 
can occur through
the accretion of outside material from mergers (M$_{\rm *,M}(t)$), and
through star formation.  We write the increase due to star formation as
$\int \psi \times \delta t$, where $\psi$ is the star formation rate within
the galaxy as a function of time (\S 3.2).  

We then write the change of stellar mass as a function of time, M$_{*}(t)$,
as

\begin{equation}
{\rm M_{\rm *}}(t) = {\rm M_{*} (0)} + {\rm M_{*, M}}(t) + \int \psi  \times \delta t  - {\rm M_{*,recy}},
\end{equation}

\noindent where ${\rm M}_{*}(0)$ is the initial galaxy stellar mass at $t=0$, 
and  ${\rm M_{*,recy}}$ is the
amount of stellar mass in stars which is converted back into a gaseous form due to
stellar evolution (\S 3.2).  

We further express how the cold gas mass (M$_{\rm g}$) changes in a similar way, 
although the cold gas mass can increase or decrease with time depending on
the star formation and gas accretion rate.

\begin{equation}
{\rm M_{\rm g}}(t) = {\rm M_{g} (0)} + {\rm M_{g, M}}(t) + {\rm M_{g, A}}(t) - \int \psi \times \delta t  + {\rm \epsilon M_{*,recy}}
\end{equation}

\noindent where in this case we have a term for the amount of cold gas brought
in from mergers, ${\rm M_{g, M}}(t)$, the amount from accretion of gas without
other galaxies\footnote{Note that this this is not strictly true, as this gas in 
principle could arise from extreme minor mergers, those with mass ratios which
are less to 1:100 and which we are not sensitive to in our observations (Bluck
et al. 2012).  However, the total amount of gas needed implies that these 
extreme minor mergers would vastly outnumber the visible minor mergers at a 
level much higher than any reasonable extrapolation.}, 
or cold gas accretion, ${\rm M_{g, A}}(t)$, as well as the amount of gas
returned to the interstellar medium from stellar evolution processes. Furthermore the
ongoing star formation within this fiducial system will lower the amount of
cold gas over time by the amount $-\int \psi \times \delta t$.  Some of the
mass created in star formation will be returned to the ISM (M$_{\rm *,recy}$),
and only a fraction $\epsilon = 0.001$ of that mass
can be reprocessed into stars, as most of the gas is shock
heated and thus cannot be used again in star formation in a time scale of
a few Gyr (Bruzual \& Charlot 2003).  We
carried out new calculations with different star formation histories showing
that even with a higher $\epsilon$ not much stellar mass is created from
this new gas.  

One of the observations of our sample which we described in \S 3.1 is that
the gas mass fraction, M$_{\rm g}$/M$_{*}$ is roughly constant over the
interval $1.5 < z < 3$ (e.g., Erb et al. 2006; Genzel et al. 2010).    In this 
case, we can write

\begin{equation}
\frac{\rm M_{g}(t)}{\rm M_{*}(t)} \sim \frac{{\rm M_{g}}(0)}{{\rm M_{*}}(0)},
\end{equation}

\noindent which is the case in the redshift range we examine for our
co-moving constant merger adjusted number density.  Note that
by using this condition the actual value of the gas to stellar mass
ratios is immaterial for the following derivation.  Thus, if we use
a different relation between star formation surface density and gas surface
density we
would obtain the same results for the basic equations, although the
final numerical result may differ (\S 4.2).  We later examine how
our net results would change if there was an evolution such that
the gas mass surface density declined at lower redshifts (\S 4.3).  

Using this measurement, we then write the amount of gas mass at a later
time $t$, M$_{\rm g}(t)$ by using equation 8.  We then expressing the 
stellar mass evolution
with time by using equation 6 to replace M$_{*}(t)$ from equation 8
to obtain

\begin{equation}
{\rm M_{\rm g}}(t) = {\rm M_{g}}(0) \times \left(1 + \frac{\rm M_{*,M}(t)}{\rm M_{*}(0)} + \frac{\int \psi \times \delta t - {\rm M_{*,recy}}}{\rm M_{*}(0)} \right),
\end{equation}

\noindent where the gas mass evolution is expressed in terms of the initial gas mass
and as a function of how much relative mass is added to a galaxy over time due 
to the increase from star formation and the increase from mergers.  These are
much easier to determine than the corresponding absolute amounts (\S 3).

Therefore using equation (9), and the results presented in \S 3,
we can write the total gas mass after a time
$\delta t$ with respect to the initial gas mass, M$_{g}(0)$ as

\begin{equation}
{\rm M_{g}}(t) \sim (2.56\pm0.21) \times {\rm M_{g}}(0),
\end{equation}

\noindent such that the total gas mass within our galaxies at $z \sim 1.5$ must
be on average $\sim$2.56 times larger than the gas mass at the start of our redshift
epoch near $z \sim 3$ due to the increase in stellar mass. 

If we then equate this to equation (7) we can calculate that
the accreted (or extra) gas mass is given by:

\begin{equation}
{\rm M_{\rm g, A}}(t) = (1.56\pm0.21) \times {\rm M_{g}}(0) + \int \psi \times \delta t -  \epsilon {\rm M_{*, recy}} -  {\rm M_{g, M}}(t).
\end{equation}

\noindent If we divide this equation by the average initial stellar mass M$_{*}(0)$ we get:

\begin{equation}
\frac{{\rm M_{\rm g, A}}(t)}{\rm M_{*}(0)} = (1.56\pm0.21) \times \frac{{\rm M_{g}}(0)}{\rm M_{*}(0)} + \frac{\int \psi \times \delta t - \epsilon M_{\rm *, recy}}{\rm M_{*}(0)} -  \frac{{\rm M_{g, M}}(t)}{\rm M_{*}(0)}.
\end{equation}

\noindent Therefore by  knowing the initial average gas mass fraction, the fraction of 
stellar mass increase in the form of stars, and the gas mass fraction brought in
through merging, we can determine the amount of gas mass, relative to our
average galaxy's initial stellar mass, brought in through `pure' accretion events, i.e., not
gas brought in with merging galaxies.  

\subsection{Gas Accretion Mass Fraction}

As we discuss in \S 3.1 we find that the initial gas mass fraction at $z = 3$ is 
f$_{\rm g} = 0.17\pm0.06, \sigma = 0.19$, and that the fraction of stellar
mass formed in star formation after accounting for stellar evolution is $= 1.00\pm0.15$ (\S 3.2),
and the amount added from mergers is M$_{\rm *,M}$/M$_{*}(0) = 0.56\pm0.15$ (\S 3.3).
The most difficult value to determine is the ratio of the cold gas mass 
brought in from mergers to the total amount of initial stellar mass 
(M$_{\rm g,M}(t)$/M$_{*}(0)$).   However, we are able to make
a measurement of this based on the amount of mass accreted from mergers
(M$_{\rm g, M}$) and the mass weighted gas mass fraction for these systems
f$_{\rm g}$ from \S 3.3.

We now use the above values, the results from \S 3, and equation (12) to calculate the
amount of gas accreted as a fraction of the initial stellar mass.
Putting these together, we conclude that 

\begin{equation}
\frac{{\rm M}_{\rm g, A}}{{\rm M}_{*}(0)} = 0.70 \pm 0.22.  
\end{equation}

\noindent This implies that a roughly constant gas mass fraction 
seen in the observations of these distant galaxies (\S 3.1) reveals that on order
the entire initial stellar mass of a massive galaxy is added
over time, outside of mergers, to form stars during $1.5 < z < 3$.  This
equates to an absolute amount of gas accreted as $(1.3 \pm 0.41) \times 10^{11}$ \solm
over 2.16 Gyr,  or at an average rate of
gas accretion of 

\begin{equation}
\frac{{\rm dM}_{\rm g, A}(t)}{\rm dt} = {\rm \dot{M}}_{\rm g, A} = (61 \pm 19)\, $\solm$ {\rm yr}^{-1}, 
\end{equation}

\noindent needed to produce the star formation we observe.   This is the amount of accreted
gas which is directly used to form stars within these massive galaxies. This is however the net amount of
gas which is accreted.  What we have not considered yet is the
effects of outflows from galaxies which will remove some fraction of the gas which is accreted.
The result of these outflows is that the gross amount of
gas actually accreted is the combination of the outflowing gas as well as the gas used
in star formation. This in principle could easily double the amount of gas mass
needed to be accreted from the IGM (e.g., Faucher-Giguere et al. 2011).

It is now well established that outflows from galaxies, particularly star forming galaxies, 
are present in observations (e.g., Heckman et al. 2000; Pettini et al. 2000; Weiner et al. 2009).  These papers show that the velocity of the outflow is 
proportional to the star formation rate (SFR) such that V$\sim$ SFR$^{0.35}$, where there is
a mild increase in the velocity of the outflow with higher star formation rate.

Weiner et al. (2009) measured the velocity outflows and rates for galaxies up to $z \sim 1.5$ 
in the DEEP2 survey.  They calculate that the mass outflow rate, ${\rm \dot{M}}_{\rm outflow}$ can be 
expressed as,

$${\rm \dot{M}_{\rm outflow}} = 22 M_{\odot}\, {\rm yr}^{-1} \left(\frac{\rm N_{H}}{10^{20} {\rm cm^{-2}}} \right) \left(\frac{\rm R}{5 {\rm kpc}}\right) \left(\frac{\rm v}{300 {\rm km\, s^{-1}}}\right).$$

\noindent Where in this equation N$_{\rm H}$ is the column density of gas, estimated by
Weiner et al. (2009) for massive galaxies to be N$_{\rm H}$ = 1.3 $\times 10^{20}$ cm$^{-2}$,
R is the radius of an expanding shell model of gas, and v is the velocity of outflows.    For our star forming galaxies
we find a velocity average from Weiner et al. (2009) just over 300 km s$^{-1}$. Combining this
with typical Petrosian sizes for these systems, as a measure of the minimum covering fraction,
we obtain an outflow rate of ${\rm \dot{M}_{outflow}} = 35$
\solm year$^{-1}$ which should be added to the net gas inflow rate that is converted into stars.  
Therefore we obtain a gross inflow rate which is 

$${\rm \dot{M}_{\rm acc} = \dot{M}_{\rm outflow}  + \dot{M}_{\rm g,A}} = 96\, {\rm M}_{\odot}\, {\rm yr}^{-1}, $$

\noindent In the next section we evaluate the uncertainties
within this measurement, and in the section after give the implications of
this result.

\subsection{Uncertainties}

There are systematic issues that could affect these results, some of 
which we examine in this section.
One of these is that the  the star formation surface density-gas surface density
relation (Schmidt-Kennicutt law) may evolve (e.g., Daddi et al. 2010).  Since
starbursts are more efficient, the initial gas mass could in principle be 
lower than
what we are measuring. If the normal Schmidt-Kennicutt law then applied for the lower 
redshift galaxies in our sample the net result would be an {\em increase} in 
the gas mass
to stellar mass ratio and equation 8 no longer strictly holds.  Likewise, it may be 
possible that we are measuring the gas masses incorrectly too low at our 
highest redshift point for some reason and that there is actually a decrease 
with time in the gas mass fraction.
We examine both of these scenarios quantitatively by considering how our 
relations would change if we use a gas mass surface density relation of the form:

\begin{equation}
\frac{\rm M_{g}(t)}{\rm M_{*}(t)} \sim \kappa(t) \times \frac{{\rm M_{g}}(0)}{{\rm M_{*}}(0)},
\end{equation}

\noindent where $\kappa (t)$ is the relative change in the ratio between the gas
to stellar mass at some lower redshift than at the initial time at $z \sim 3$.
In this case, the relative gas mass accretion can be written as:

\begin{equation}
\frac{{\rm M_{\rm g, A}}(t)}{\rm M_{*}(0)} = \kappa \times 1.56 \times \frac{{\rm M_{g}}(0)}{\rm M_{*}(0)} + \frac{\int \psi \times \delta t}{\rm M_{*}(0)} -  \frac{{\rm M_{g, M}}(t)}{\rm M_{*}(0)} -  \frac{\epsilon {\rm M_{*,recy}}}{\rm M_{*}(0)}.
\end{equation}

\noindent When using this equation with the value of $\kappa = 0.5$, where the gas mass fraction
has dropped by a half, we find that the gas accretion mass fraction drops to  
M$_{\rm g, A}$/M$_{*} = 0.56 \pm 0.22$.  If we take the limiting case where all of the
gas is exhausted from $z = 3$ to $z = 1.5$, where $\kappa = 0$, we would find a lower
limit accreted gas mass
fraction of M$_{\rm g, A}$/M$_{*} = 0.43 \pm 0.22$.   

In general, if the gas mass 
fraction decreases with time, with perhaps an evolving form of the relation between 
gas mass surface density and
star formation rate surface density, this would result in a slightly lowered 
derived gas mass
accretion.  For the more likely case of less efficient star forming systems
over time, the net gas mass fraction would increase even more than our initial calculation, given 
that in this case $\kappa > 1$.   However, most of the constraint on this measurement comes from the fact
that there is a high ongoing star formation rate during this epoch, and the gas
to fuel this star formation is not initially present within the galaxy, nor
carried in through mergers.

If we directly consider
the more efficient Schmidt-Kennicutt law, as proposed by Daddi et al. (2010) for
star bursting galaxies to apply throughout our redshift range, we would then have a lower
gas mass fraction at both the start and the end of our evolution.  As above, we would
still have to account for the star formation present within these systems, and this
is the driving observation behind our calculation of a high gas mass accretion rate. 
However it is important to note that our galaxies are in the regime where they follow 
the standard law (Daddi et al. 2010), where the more efficient starburst are
found within sub-mm and ULIRG galaxies (Figure~2).  

One remaining issue is that our star formation rates maybe too high,
however, others have found very similar star formation rates or higher, for
similar galaxies (e.g., Daddi et al. 2007).   The star formation rates we measure
are consistent in the ultraviolet and using {\em Herschel} far-infrared imaging
(Hilton et al. 2012), and if anything our values are at the lower end of the various star
formation measurements (Bauer et al. 2011).

\subsection{Implications}

We present in this paper an analysis of the amount of cold gas mass which is 
likely accreted into massive galaxies with stellar masses M$_{*} >$ \mass 
at $1.5 < z < 3$.  This amount of accreted gas is necessary 
to account for the observed star formation rate between redshifts $1.5 < z < 3$, 
which cannot be accounted for by gas brought in through minor+major mergers 
(Bluck et al. 2012).  Our overall result is that we find a basic
accretion rate of ${\rm \dot{M}}_{\rm g, A}$ = (96 $\pm$ 19) \solm yr$^{-1}$ of 
cold gas needed from the IGM to account for this star formation.    This
accounts for 66$\pm 20$\% of the gas brought into these galaxies and ultimately
responsible for 49$\pm 20$\% of the stellar mass assembly produced between $1.5 < z < 3$.
In addition to this we find that 25$\pm 10$\% of the stellar mass accreted is from
mergers, while the remaining $\sim 25$\% is from star formation produced in
gas accretion associated with mergers.  Therefore the bulk of the formation of massive
galaxies at $1 < z < 3$ is due to conversion of baryons into gas as observed
in the star formation rate.

Overall this implies that gas accretion into massive galaxies
 at early epochs is a major formation method, and may dominate over mergers as a 
formation process for new stars.  We also note that we are not certain what is the origin of
this gas accretion.  It could originate from cold filaments, or from gas cooling
from the halo itself (e.g., Keres et al. 2005; Ocvirk et al. 2008), and we are not able to address 
this directly.

Papovich et al. (2011) measure the total 
gas accretion rate at similar redshifts, which
includes all methods that bring in gas.  They were not able to distinguish between the gas
brought in through mergers, and that brought in through gas accretion.  
Their results are similar to ours, finding a higher accretion rate, but that 
the total accretion is similar to what we find here, or slightly higher as it includes
gas brought in through minor mergers.  
We note that we are only able to probe mergers for massive 
galaxies down to M$_{*} = 10^{9.5}$ \solm, and it is possible that the gas accretion we
find originates from galaxies with stellar masses lower than this limit. Furthermore, 
we are not able to say whether the accreted gas originates in clumps of gas into the 
galaxy or in a smooth continuous
mode.  We also cannot rule out that some of this gas is originating from the
outer halo after cooling. 
Our result is an integrated average of gas accretion through these modes
over the time-spanned by redshifts $1.5 < z < 3$.  

There are several implications from these results.  Our finding may
explain how some massive galaxies in the very local universe, outside the Local
Group but within 8 Mpc, can have
almost no bulges (e.g., Kormendy et al. 2010).  Gas accretion is also one way to solve the 
G-dwarf problem of having too
many metal rich stars in the solar neighbourhood (e.g., Larson et al. 1974), and may relate
to the gas accretion we see in our own galaxy (Blitz et al. 1999). 

Overall our result of finding an accretion rate of gas of $\dot{M}_{\rm acc}$ =
(96 $\pm$ 19) \solm yr$^{-1}$  for the most
massive galaxies at $1.5 < z < 3$ is roughly consistent with theoretical calculations
which predict a similar amount of gas accretion (Murali et al. 2002; van den Bosch 2002;
Keres et al. 2005;  Dekel et al. 2009a,b).
Early simulations
by Murali et al. (2002) predict a gas accretion rate of $\dot{M}_{\rm g, A}$ $\sim 40$ \solm yr$^{-1}$,
while more recent work suggests higher rates of $\dot{M}_{\rm g, A}$ $\sim 100$ \solm yr$^{-1}$
(e.g., Dekel et al. 2009a,b; Faucher-Giguere et al. 2011). Our results are in
general agreement with these models, although we are only examining a narrow
redshift and stellar mass interval within this paper.
 
Our results from this work, and previous papers examining the issue of stellar mass 
formation  modes (e.g., Conselice 2006; Bluck et al. 2012) is similar to what is predicted
in several studies using N-body simulations. This includes Genel et al. (2010), who
find through simulations that about 60\% of the dark matter mass in galaxies is put into 
place through 
merging, with at least 40\% of the mass coming from accretion.  This is however, for
the formation of the dark matter halo and not the baryons or stars.
Using hydrodynamical  galaxy formation models, Murali 
et al. (2002) find that for galaxies brighter than about a 
fourth of L$^{*}$, the characteristic luminosity, gas accretion dominates over merging as a 
mechanism, whereby gas accretion accounts for $\sim 75$\% of the stellar mass build up 
at $z \sim 2$.  While others, e.g., Stewart et al. (2008) find that over half of the mass 
in galaxies is formed in mergers, while others such as Angulo \& White (2010) find that 
nearly all the mass in galaxies is put into place via mergers.    Finally Keres
et al. (2009) find that the minor+merger mass assembly is comparable to the formation
due to gas accretion.   There is therefore considerable disagreement between these
various models about which mode is the dominant. We find that the gas accretion and
mergers are about as equally important.  
Future studies will have to probe deeper to obtain the
merger history for lower mass systems to carry out similar calculations as these, and probe
this evolution in different environments.

\section{Summary}

We present in this paper a study of the cold gas mass densities for massive
galaxies with M$_{*} > 10^{11}$~\solm at redshifts of $1.5 < z < 3$.  While we
do not directly detect the accretion of gas within our galaxies, we are able
to make a circumstantial finding for its existence.   Our conclusions
are as follows:

I. Utilizing our measured star formation rates and galaxy sizes we find
a roughly constant cold gas mass to stellar mass fraction 
for this sample across the redshift
range of $1.5 < z < 3$.

II.  We utilize the star forming and merging properties of
these galaxies from previous work in Bauer et al. (2011) and Bluck et al.
(2012) to measure the mass budget of our sample of massive galaxies, finding
that $\dot{M}_{\rm acc}$ = (96 $\pm$ 19) \solm yr$^{-1}$ of gas is needed to 
sustain the
star formation rate outside of gas brought in via mergers.

III. We derive based
on these values that cold gas accretion from the intergalactic medium, or alternatively 
very minor galaxy mergers with mass ratios lower than 1:100 (or 1:10 in ratios
of baryons), accounts for 49$\pm$20\%
of the baryonic matter added to galaxies from $1.5 < z < 3$.  

This amount of gas mass added from accretion is larger than the amount of gas added due
to the merger process (both minor and major) (e.g., Conselice 2006; Bluck
et al. 2012) and is largely in agreement with models
which predict on the order of 100-200 \solm yr$^{-1}$ added from cold gas accretion
(e.g., Dekel et al. 2009a,b).    Gas accretion is therefore the major
method for producing star formation within massive galaxies between redshifts
$1.5 < z < 3$.  

 Further work with e.g., the CANDELS survey will allow us to
carry out this measurement for lower mass galaxies where the mode of
formation could be significantly different than the more massive systems
(e.g., Dekel et al. 2009a,b).  Also, we are examining in this paper 
massive systems in the last throws of their formation at $z < 3$. Investigating
the ratio of formation due to mergers and gas accretion at $z > 3$ will reveal
how the first epochs of these massive galaxies were formed.  This however
will require observations from JWST and the ELTs.

We thank the GNS team, particularly Fernando Buitrago and Amanda Bauer,
for their contributions to this survey and the previous published work
utilised here.  We thank the referee for a report that improved
this paper significantly. 
The data and catalogs as used in the GNS survey are online at:
{\bf http://www.nottingham.ac.uk/astronomy/gns/}.
The GNS is financially supported by STFC and the Leverhulme Trust.
Support was also provided by NASA/STScI grant HST-GO11082.

\appendix

\label{lastpage}

\end{document}